\documentclass{iopart} 

\usepackage{enumerate}

\usepackage[T1]{fontenc}

\expandafter\let\csname equation*\endcsname\relax
\expandafter\let\csname endequation*\endcsname\relax
\usepackage{amsmath, amsthm, amssymb, amsfonts}

\usepackage{graphicx}
\usepackage[colorlinks=true,linkcolor=blue,citecolor=blue,filecolor=cyan,urlcolor=blue,breaklinks=false]{hyperref}

\usepackage{cite}

\newcommand{\mrd}{\mathrm d}
\newcommand{\mre}{\mathrm e}
\newcommand{\pstat}{p^{\mathrm s}}
\newcommand{\jstat}{J^{\mathrm s}}
\newcommand{\jstatc}{\mathcal{J}^\mathrm{s}}
\newcommand{\mean}[1]{\left\langle #1 \right\rangle}
\newcommand{\sgn}{\operatorname{sgn}}
\newcommand{\arsinh}{\operatorname{arsinh}}
\renewcommand{\vec}[1]{\boldsymbol{#1}}

\newcommand{\mcJ}{\mathcal{J}}

\begin{document}
\title{Affinity- and topology-dependent bound on current fluctuations}
\author{Patrick Pietzonka$^1$, Andre C. Barato$^2$, and Udo Seifert$^1$}

\address{$^1$ II. Institut f\"ur Theoretische Physik, Universit\"at Stuttgart, 70550 Stuttgart, Germany}
\address{$^2$ Max Planck Institute for the Physics of Complex Systems,
  N\"othnitzer Stra\ss e 38, 01187 Dresden, Germany}

\begin{abstract}
  We provide a proof of a recently conjectured universal bound on
  current fluctuations in Markovian processes. This bound establishes
  a link between the fluctuations of an individual observable current,
  the cycle affinities driving the system into a non-equilibrium
  steady state, and the topology of the network. The proof is based on
  a decomposition of the network into independent cycles with both
  positive affinity and positive stationary cycle current. This
  formalism allows for a refinement of the bound for systems in
  equilibrium or with locally vanishing affinities.\\[1\baselineskip]
\end{abstract}
\vspace{1cm}

\paragraph{Introduction.} A comprehensive mathematical description of systems driven into a
non-equilibrium steady state (NESS) is an open challenge in statistical
physics. So far, the arguably most prominent universal property of these
systems is the fluctuation theorem, establishing a symmetry between the
probability of current fluctuations and the corresponding sign-reversed
fluctuations \cite{lebo99,andr07c,pole14,seif12}. Recently, we have proposed bounds
on the probability of current fluctuations, which hold universally in systems
that can be represented in terms of continuous-time Markov networks
\cite{piet15}. These bounds are formulated within the mathematical framework
of large deviation theory \cite{ellis,touc09}. This theory describes the decay of the
probability of current fluctuations with time $t$ as an exponential law
\begin{equation}
  p(J,t)\sim\mre^{-tI(J)},
  \label{eq:ldfdef}
\end{equation}
where $J$ denotes a generic fluctuating current, e.g., a cycle current, a current along an
individual transition, the current associated with the entropy
production or even a vector of several such
currents \cite{piet15}. The non-negative and convex function $I(J)$ is called a \textit{rate function} or
\textit{large deviation function}. A dual description of the exponential law
\eqref{eq:ldfdef} is given in terms of the scaled cumulant generating function
\begin{equation}
  \lambda(z)\equiv\lim_{t\to\infty}\frac{1}{t}\log\mean{\mre^{zJt}}\equiv\lim_{t\to\infty}\frac{1}{t}\log\int\mrd J\,p(J,t)\,\mre^{zJt}=\max_J[zJ-I(J)],
  \label{eq:scgfdef}
\end{equation}
where the integral runs over all possible realizations of the current
$J$. The function $\lambda(z)$ is the Legendre-Fenchel transform of
$I(J)$. In Ref.~\cite{piet15}, we have presented several upper bounds
on $I(J)$ and respective lower bounds on $\lambda(z)$. In particular, a
parabolic bound depends only on the average entropy production of the
system. Recently, this bound was proven by Gingrich \etal
\cite{ging16}. Building partially on the ideas in their proof, we
prove a second, stronger bound that requires additional information on
the affinities and the topology of the Markov network
\cite{piet15}. This bound is saturated for an asymmetric random walk
driven by a single affinity, for which $\lambda(z)$ has the shape of a
hyperbolic cosine. Remarkably, this shape persists as a bound for
processes on finite networks with arbitrarily complex topology and
non-uniform transition rates. Since by construction the bounds are
saturated for the most likely, stationary value $J=\jstat$ with
$I(\jstat)=0$ (or, respectively, for $z=0$ with $\lambda(0)=0$ and
$\lambda'(0)=\jstat$), they imply particularly strong bounds on the
typical Gaussian fluctuations around $\jstat$, which have been
previously conjectured in terms of the thermodynamic uncertainty
relation \cite{bara15} and an affinity dependent bound on the Fano
factor \cite{bara15a}. In these forms, the bounds can be
experimentally applied to, e.g., enzymatic reaction networks or small
electronic circuits.

\paragraph{Setup.} We consider a Markov process on a finite network of $N$ states
$\left\{i\right\}$ with transition rates $k_{ij}$ from state $i$ to state $j$
and assume micro-reversibility, i.e., $k_{ij}>0$ implies $k_{ji}>0$. The
stationary probability distribution $\pstat_i$, satisfying
\begin{equation}
  \sum_{j}(\pstat_i k_{ij}-\pstat_{j} k_{ji})=0,
\end{equation}
gives rise to the stationary currents
\begin{equation}
 \jstat_{ij}\equiv \pstat_i k_{ij}-\pstat_j k_{ji}=-\jstat_{ji}.
\label{eq:jstatdef}
\end{equation}
These stationary currents are the expectation values of the \textit{fluctuating}
currents $J_{ij}$, defined as the net number of transitions in a
stochastic trajectory along the edge
$(i,j)$ divided by the observation time $t$.
An individual transition $i\to
j$ contributes \cite{seif12}
\begin{equation}
  f_{ij}\equiv\log\frac{\pstat_ik_{ij}}{\pstat_j k_{ji}}
\label{eq:fdef}
\end{equation}
to the entropy production in a NESS,
leading to the average total entropy production 
\begin{equation}
  \sigma\equiv\sum_{i<j}\jstat_{ij}f_{ij}.
  \label{eq:sm}
\end{equation}
We denote by $\sum_{i<j}$ the sum over those edges
$(i,j)$ in the network graph for which transitions are possible,
i.e., for which $k_{ij}>0$ and $k_{ji}>0$.

The decomposition of networks into cycles is an important concept
providing a link between (bio-)physical properties of a system and its
description as a stochastic process \cite{schn76,hill,qian07}.
A cycle $\mathcal{C}_\alpha$ of length $n_\alpha\geq 3$ (smaller cycles are
not relevant for our discussion) is defined as a directed, self-avoiding, closed path
$[\ell(1)\to \ell(2)\to\dots\to \ell(n_\alpha)\to \ell(1)]$ along edges of the
network with $k_{\ell(n)\ell(n+1)}>0$.  We define the directed adjacency
matrix $\vec{\chi}^\alpha=(\chi_{ij}^\alpha)$ of such a cycle as
\begin{equation}
  \chi_{ij}^\alpha\equiv\sum_{n=1}^{n_\alpha}\delta_{\ell(n),i}\delta_{\ell(n+1),j}-\delta_{\ell(n),j}\delta_{\ell(n+1),i},
\label{eq:chi}
\end{equation}
which is $+1$ for edges $(i,j)$ where $\mathcal{C}_\alpha$ passes in forward
direction, $-1$ for the backward direction and zero otherwise.
A set of cycles $\left\{\mathcal{C}_\alpha\right\}$ is called complete if any
set of currents $\{J_{ij}\}$ along the edges of the network
consistent with Kirchhoff's law
\begin{equation}
  \sum_{j}J_{ij}=0\qquad\textup{for all $i$}
  \label{eq:kirchhoff}
\end{equation}
can be decomposed into a set of cycle currents
$\left\{\mcJ_\alpha\right\}$ such that
\begin{equation}
  J_{ij}=\sum_\alpha\mcJ_\alpha\chi^\alpha_{ij}.
  \label{eq:currdecomp}
\end{equation}
In particular, the stationary currents $\jstat_{ij}$ can be decomposed into
stationary cycle currents $\jstatc_\alpha$. The affinity
$\mathcal{A}_\alpha$ of a cycle is given by 
\begin{equation}
  \mathcal{A}_\alpha\equiv\sum_{i<j}\chi_{ij}^\alpha\log\frac{k_{ij}}{k_{ji}}=\sum_{i<j}\chi_{ij}^\alpha
  f_{ij}=\sum_{(i,j)\in\mathcal{C}_\alpha}f_{ij},
\label{eq:affdef}
\end{equation}
which  can be used to write the average entropy production \eqref{eq:sm} as
\begin{equation}
  \sigma=\sum_\alpha\jstatc_\alpha\mathcal{A}_\alpha.
  \label{eq:smc}
\end{equation}

\paragraph{Main result.} With the above definitions at hand, we can now state the central result
proven in this Letter. The scaled cumulant generating function 
\begin{equation}
  \lambda(\vec
  z)\equiv\lim_{t\to\infty}\frac{1}{t}\log\mean{\mre^{\vec
      z\cdot\vec{\mathcal J}t}} 
\end{equation}
for the fluctuations of the vector $\vec{\mathcal J}=(\mathcal
J_\alpha)$ of cycle currents is bounded from below by
\begin{equation}
  \label{eq:mainresultshort}
  \lambda(\vec z)\geq \sigma\frac{\cosh[(\vec z\cdot
    \vec{\jstatc}/\sigma+1/2)\mathcal{A}^*/n^*]-\cosh[\mathcal{A}^*/(2n^*)]}{(\mathcal{A}^*/n^*)\sinh\mathcal[\mathcal{A}^*/(2n^*)]},
\end{equation}
where $\mathcal A^*/n^*$ is defined as the smallest positive value for
the affinity per cycle length over all cycles in the network. In a
scalar version, this bound implies for the scaled cumulant generating
function \eqref{eq:scgfdef} of any generic current $J$, which can be
written as an expansion in the cycle currents, the bound
\begin{equation}
\label{eq:mainresultscalar}
   \lambda(z)\geq \sigma\frac{\cosh[( z
    {\jstat}/\sigma+1/2)\mathcal{A}^*/n^*]-\cosh[\mathcal{A}^*/(2n^*)]}{(\mathcal{A}^*/n^*)\sinh\mathcal[\mathcal{A}^*/(2n^*)]},
\end{equation}
where $\jstat$ is the stationary value of $J$. In particular, for the
entropy current $\sum_\alpha\mcJ_\alpha\mathcal{A}_\alpha$ we have
$\jstat=\sigma$, which further simplifies \eqref{eq:mainresultscalar}.
Based on strong numerical evidence we have conjectured the relation \eqref{eq:mainresultshort}
in Ref.~\cite{piet15}. Furthermore, Eq.~\eqref{eq:mainresultshort}
generalizes a bound on the Fano factor in enzyme kinetics that has
been conjectured in Ref.~\cite{bara15a}.

\paragraph{Uniform cycle decomposition.} For multicyclic networks, the
choice of a minimal complete set of cycles is not unique. We make use
of this ambiguity to expand the stationary current $\jstat_{ij}$ in a
specific set of cycles $\{\mathcal{C}_\beta\}$.  This set is
constructed in a way that every cycle $\mathcal{C}_\beta$ contributing
to the stationary current, i.e., for which $\jstatc_\beta\neq 0$,
satisfies the following properties, which will turn out to be
essential for our proof of the affinity dependent bound \eqref{eq:mainresultshort}:
\begin{enumerate}[(i)]
\item The edges of every cycle are aligned with the stationary current, i.e., 
  \begin{equation}
    \sgn \chi^\beta_{ij}=\sgn \jstat_{ij}=\sgn f_{ij}
    \label{eq:uniformdef}
  \end{equation}
  for all $i$, $j$ and $\beta$. We call such cycles \textit{uniform} with
  respect to $\jstat_{ij}$. The
  equality with $\sgn f_{ij}$ follows directly from the definitions
  \eqref{eq:jstatdef} and \eqref{eq:fdef} and shows via Eq.~\eqref{eq:affdef}
  that uniform cycles have positive affinity.
\item All stationary cycle currents are strictly positive, $\jstatc_\beta>0$ for all $\beta$.
\end{enumerate}
It can be easily checked that an arbitrary cycle decomposition, for
example the decomposition into fundamental cycles by Schnakenberg
\cite{schn76}, does not necessarily meet these conditions. A
construction of a cycle decomposition that satisfies at least
condition (ii) was introduced by J.~MacQueen \cite{macq81}, its physical
relevance and proof is also discussed in Ref.~\cite{alta12a}.
This decomposition, however, refers only to networks with
irreversible transitions. The network of our present setup with
reversible transitions could be mapped on such a network by replacing
every edge by two irreversible links with antiparallel direction, yet
this procedure does not necessarily lead to a decomposition satisfying
condition (i).

\begin{figure}
  \centering
  \includegraphics{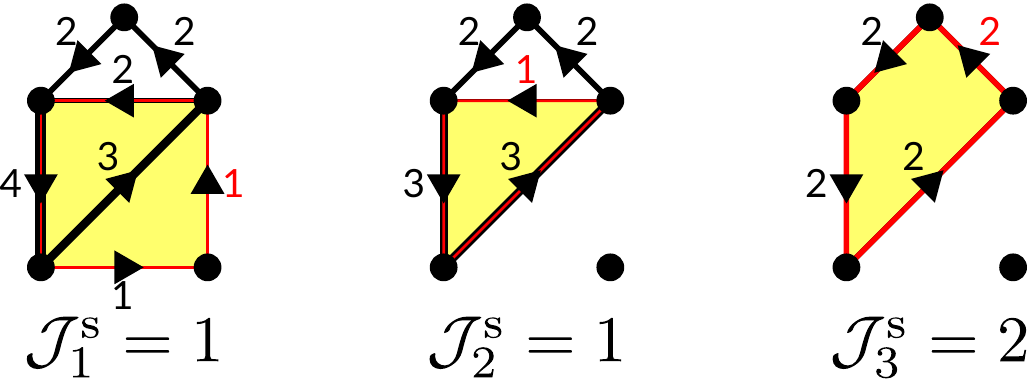}
  \caption{Example for a decomposition of a 5-state network into a complete
    set of three independent uniform cycles. Edges are labeled by the
    strength of the currents $J_{ij}^{(\beta)}$, whose direction is
    marked by arrows. For each of the three iteration steps, an edge with
    minimal current is labeled in red and the uniform cycle
    $\mathcal{C}_\beta$ including this edge is shown as a yellow polygon with
    red border.}
  \label{fig:monocycle}
\end{figure}

\begin{figure}
  \centering
  a)\raisebox{-\height}{\includegraphics[scale=0.7]{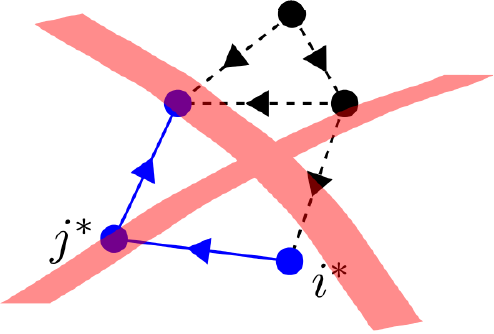}}
  b)\raisebox{-\height}{\includegraphics[scale=0.7]{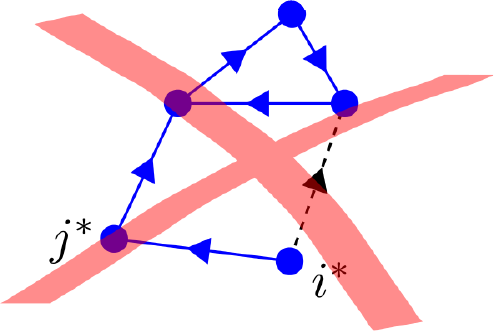}}
  c)\raisebox{-\height}{\includegraphics[scale=0.7]{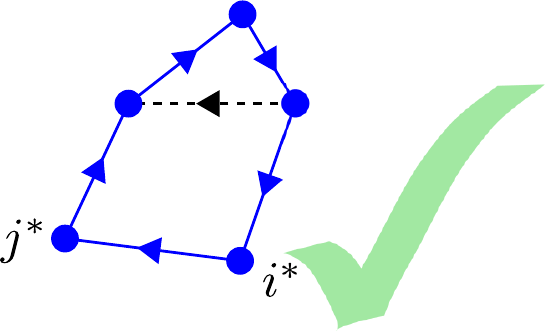}}
  \caption{Illustration of a uniform cycle
    $\mathcal{C}_\beta$ starting with the edge $(i^*,j^*)$. The direction of
    $J_{ij}^{(\beta)}$ is indicated by the arrows, the (attempted) cycle is
    shown as solid blue lines. Neither dead ends (a) nor ``dead-cycles'' (b) are
    consistent with Kirchhoff's law \eqref{eq:kirchhoff}.}
  \label{fig:kirchhoff}
\end{figure}

Here, we
present a variation of the algorithm of MacQueen for networks with
genuinely reversible transitions, which generates a set of cycles
satisfying both conditions (i) and (ii), as exemplified in
Fig.~\ref{fig:monocycle}.
\begin{itemize}
\item Initialization. Set $J_{ij}^{(1)}\equiv\jstat_{ij}$.
\item Iteration over $\beta\geq 1$. Locate an edge $(i^*,j^*)$ with minimal
  positive current $J_{i^*j^*}^{(\beta)}$ and set
  \begin{equation}
    \jstatc_\beta\equiv \min_{i,j|J_{ij}^{(\beta)}>0}J_{ij}^{(\beta)}=J_{i^*j^*}^{(\beta)}.
  \end{equation}
  Construct $\mathcal{C}_\beta$ as a closed self avoiding path
  starting with the edge $(i^*,j^*)$ and passing only along edges in
  the direction for which $J_{ij}^{(\beta)}>0$. This path is not
  necessarily unique, one of several such paths can be chosen freely
  as $\mathcal{C}_\beta$. Using the directed adjacency matrix
  $\chi_{ij}^\beta$ corresponding to $\mathcal{C}_\beta$, as defined
  in Eq.~\eqref{eq:chi}, set
  \begin{equation}
    J_{ij}^{(\beta+1)}=J_{ij}^{(\beta)}-\jstatc_\beta\chi^\beta_{ij}.
  \end{equation}
\item Terminate when $J_{ij}^{(\beta+1)}=0$ for all $i,j$.
\end{itemize}
By construction, the currents $J_{ij}^{(\beta)}$ satisfy
Kirchhoff's law \eqref{eq:kirchhoff} for every $\beta$. As a
consequence, it is always possible to construct an appropriate uniform
cycle $\mathcal C_\beta$ following the direction of
$J_{ij}^{(\beta)}$: For every node where there is a way in, there must
also be a way out, which rules out dead ends (see
Fig.~\ref{fig:kirchhoff}a). Similarly, it is not possible to run into
a cycle without exit that does not include the starting node $i^*$
(see Fig.~\ref{fig:kirchhoff}b). On a finite set of states, every self
avoiding path must reach the starting point $i^*$ again after at most
$N$ steps. Typically, there is more than one cycle starting with
$(i^*,j^*)$ meeting these conditions. Any of these cycles can be
selected as $\mathcal C_\beta$. Since we use in every iteration the
minimal current as new cycle current, the individual currents
$J^{(\beta+1)}_{ij}$ are either zero or have the same sign as
$\jstat_{ij}$. Thus, every cycle that is uniform with respect to the
currents $J^{(\beta)}_{ij}$ is also uniform with respect to $\jstat_{ij}$,
as required by condition (i). The matrices $\chi^\beta_{ij}$ are
linearly independent, since for every $\beta$ the edge $(i^*,j^*)$ is
no longer contained in all the subsequent cycles.  By construction,
the algorithm leads to the decomposition
\begin{equation}
  \jstat_{ij}=\sum_\beta\jstatc_\beta\chi^\beta_{ij}.
  \label{eq:jstatdecomp}
\end{equation}
Typically, the algorithm terminates when $\beta$ reaches the number of
fundamental cycles. It cannot terminate later since all cycles
$\mathcal{C}_\beta$ are linearly independent.  It may happen that the
algorithm terminates earlier, so that the set of cycles
$\{\mathcal C_\beta\}$ is not complete, i.e., it cannot be used to
represent arbitrary currents different from the stationary current.
In this case the set $\{\mathcal C_\beta\}$ can be completed by
adding further linear independent non-uniform cycles. The stationary
current of these cycles is always zero.

\paragraph{Proof.}
Equipped with the decomposition in uniform cycles, we can now prove
the affinity dependent bound \eqref{eq:mainresultshort} on the large
deviation function. As a starting point we use, building on
Ref.~\cite{maes08} and analogously to Ref.~\cite{ging16}, the
universally valid bound for the large deviation function of the
distribution of the vector $\vec J=(J_{ij})$ of all fluctuating currents
along edges,
\begin{equation}
  I(\vec
  J)\leq\sum_{i<j}\left[J_{ij}\left(\arsinh\frac{J_{ij}}{a_{ij}}-\arsinh\frac{\jstat_{ij}}{a_{ij}}\right)-\left(\sqrt{a_{ij}^2+J_{ij}^2}-\sqrt{a_{ij}^2+(\jstat_{ij})^2}\right)\right],
\label{eq:masterbound}
\end{equation}
with $a_{ij}\equiv 2\sqrt{\pstat_i\pstat_j k_{ij} k_{ji}}$. This
relation holds for all values of $\vec J$ that are consistent with
Kirchhoff's law \eqref{eq:kirchhoff}, otherwise
$I(\vec{J})=\infty$. Eq.~\eqref{eq:masterbound} follows from an exact
expression for the ``level 2.5'' large deviation function
$I(\vec J,\vec p$) for the joint distribution of the fluctuating
current $\vec J$ and the fluctuating density $\vec p=(p_i)$
\cite{bara15d,maes08}.  By using the contraction principle, the large
deviation function for the currents can be expressed as
$I(\vec J)=\min_{\vec p}I(\vec J,\vec p)=I(\vec J,\vec p^*(\vec
J))$.
Using the stationary distribution $\vec\pstat$ instead of the most
likely density $\vec p^*(\vec J)$, one obtains as an upper bound on
$I(\vec J)$ the function $I(\vec J,\vec\pstat)$, which can be brought
to the form of the r.h.s. of Eq.~\eqref{eq:masterbound}. The choice
$\vec\pstat$ ensures that the bound is saturated for the most
likely current $\vec J=\vec\jstat$.

\begin{figure}
  \centering
  \includegraphics{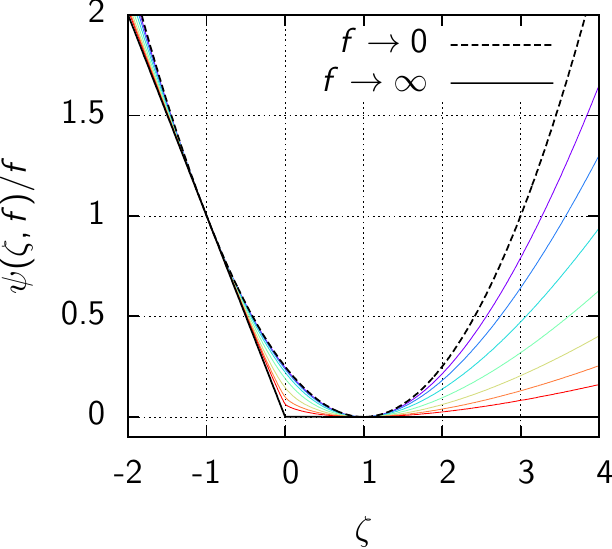}\qquad
  \includegraphics{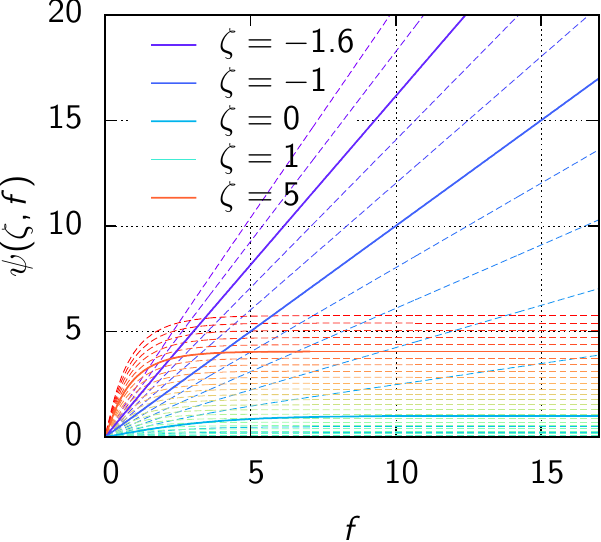}
  \caption{Left: The function $\psi(\zeta,f)/f$ as a function of
    $\zeta$ for values of $f$ ranging from 1 (violet) to $10^{1.2}$
    (red) with logarithmic spacing. For fixed $\zeta$ and $f>0$,
    $\psi(\zeta,f)/f$ decreases monotonically in $f$, as required in
    Eq.~\eqref{eq:hdomA}. The limiting curves for
    $\lim_{f\to\infty}\psi(\zeta,f)/f=(|\zeta|-\zeta)/2$ and for
    $\lim_{f\to 0}\psi(\zeta,f)/f=(\zeta-1)^2/4$ are shown in
    black. Right: The function $\psi(\zeta,f)$ as a function of $f$
    for values of $\zeta$ ranging from $-2$ (violet) to $6$
    (red). Selected values of $\zeta$ are shown as solid lines, in
    particular $\psi(-1,f)=f$ and $\psi(1,f)=0$.  Obviously,
    $\psi(\zeta,f)$ is concave for $f\geq 0$. }
  \label{fig:psiplot}
\end{figure}

At first, we assume for simplicity that the stationary current $\jstat_{ij}$
is non-zero along all edges with $k_{ij}>0$. Then, we can rewrite the bound
\eqref{eq:masterbound} as
\begin{equation}
  I(\vec{J})\leq \sum_{i<j}\jstat_{ij}\psi(J_{ij}/\jstat_{ij},f_{ij}),
  \label{eq:gingbound}
\end{equation}
where the function $\psi(\zeta,f)$ is defined as 
\begin{equation}
  \psi(\zeta,f)\equiv\zeta\arsinh(\zeta/b)-\zeta
  f/2-\sqrt{b^2+\zeta^2}+\sqrt{b^2+1}
  \label{eq:psidef}
\end{equation}
with
\begin{equation}
  b\equiv [\sinh(f/2)]^{-1}.
\end{equation}
The function $\psi(\zeta,f)$ is the large deviation function for the
current in an asymmetric random walk on a ring with uniform transition
rates, with affinity per step $f$, and with stationary current
$\jstat=1$ \cite{lebo99}. For this highly symmetric network, there is
only one possible current $J$ and $\pstat_i=p_i^*(J)=1/N$ holds
independently of $J$, so that Eq.~\eqref{eq:gingbound} becomes an
equality.  For positive $f$ and fixed $\zeta$, $\psi(\zeta,f)$ is
concave in $f$, moreover, $\psi(\zeta,f)/f$ decreases monotonically
with increasing $f$, as proven in \ref{sec:appa} and \ref{sec:appb},
respectively. These essential properties are demonstrated in
Fig.~\ref{fig:psiplot}. The Legendre transform of $\psi(\zeta,f)$ is
\begin{equation}
  \lambda(y,f)\equiv\max_\zeta[y\zeta-\psi(\zeta,f)]=[\cosh(y+f/2)-\cosh(f/2)]/\sinh(f/2).
\label{eq:legendrelambda}
\end{equation}

Using the decomposition in the uniform cycles $\{\mathcal{C}_\beta\}$
from Eq.~\eqref{eq:jstatdecomp}, for which all $f_{ij}$ along a cycle
become positive, we can write Eq.~\eqref{eq:gingbound} as
\begin{equation}
  I(\vec J)\leq\sum_{\beta}\jstatc_\beta\sum_{(i,j)\in
    \mathcal{C}_\beta}\psi(J_{ij}/\jstat_{ij},f_{ij}).
  \label{eq:beforejensen}
\end{equation}
Next, we set $J_{ij}=\xi \jstat_{ij}$ for all $i,j$. This choice simplifies the following steps and ensures
that the fluctuating currents $J_{ij}$ are consistent with Kirchhoff's
law \eqref{eq:kirchhoff}. Having ensured via the uniform cycle
decomposition that all $f_{ij}$ are positive, we can apply Jensen's
inequality to the second sum in Eq.~\eqref{eq:beforejensen}, in which
the summation index runs over the $n_\beta$ edges of the cycle
$\mathcal{C}_\beta$. Since the $\jstatc_\beta$ are positive, this
procedure leads to
\begin{equation} 
  I(\xi\vec J^\mathrm{s})\leq\sum_{\beta}\jstatc_\beta n_\beta\sum_{(i,j)\in
    \mathcal{C}_\beta}\frac{1}{n_\beta}\psi(\xi,f_{ij})\leq
  \sum_\beta\jstatc_\beta \mathcal{A}_\beta
  (n_\beta/\mathcal{A}_\beta)\,\psi(\xi,\mathcal{A}_\beta/n_\beta),
  \label{eq:jensen}
\end{equation}
where we have used Eq.~\eqref{eq:affdef}. For a coarser bound, we
identify the uniform cycle with minimal
$\mathcal{A}_\beta/n_\beta\equiv\mathcal{A}^*/n^*$, which is positive
as explained below Eq.~\eqref{eq:uniformdef}. Using the monotonic
decrease of $\psi(\zeta,f)/f$, the positivity of
$\jstatc_\beta\mathcal{A}_\beta$, and Eq.~\eqref{eq:smc}, we obtain
\begin{equation}
  I(\xi\vec{J}^\mathrm{s})\leq \sigma\,(n^*/\mathcal{A}^*)\,\psi(\xi,\mathcal{A}^*/n^*).
  \label{eq:hdomA}
\end{equation}
Finally, we adopt a more elegant formulation of the large
deviation function taking the cycle currents $\mcJ_\beta$ as argument instead of
$(J_{ij}) =\vec J$, i.e., we write 
\begin{equation}
\mathcal{I}(\vec\mcJ)\equiv
I\left(\sum_\beta\mcJ_\beta\vec\chi^\beta\right).  
\label{eq:Icycle}
\end{equation}
We thus avoid currents that are inconsistent with Kirchhoff's law, for
which the large deviation function would become infinite. The scaled
cumulant generating function corresponding to the large deviation
function \eqref{eq:Icycle} turns out to be bounded by
\begin{align}
  \lambda(\vec z)&=\max_{\vec\mcJ}\left[\sum_\beta
    z_\beta\mcJ_\beta-\mathcal{I}(\vec{\mathcal J})\right]\nonumber\\
  &\geq \max_{\vec \mcJ=\xi\vec{\mcJ}^\mathrm{s}}\left[\vec
    z\cdot\vec{\jstatc}\xi-I(\xi\vec{J}^\mathrm{s})\right]\nonumber\\
  &\geq \max_{\xi}\left[\vec
    z\cdot\vec{\jstatc}\xi-
    \sigma\,(n^*/\mathcal{A}^*)\,\psi(\xi,\mathcal{A}^*/n^*)
  \right]\nonumber\\
&=\frac{\sigma}{\mathcal{A}^*/n^*}\lambda(\vec z\cdot \vec{\jstatc}(\mathcal{A}^*/n^*)/\sigma,\mathcal{A}^*/n^*)\nonumber\\
  &=  \sigma\frac{\cosh[(\vec z\cdot
    \vec{\jstatc}/\sigma+1/2)\mathcal{A}^*/n^*]-\cosh[\mathcal{A}^*/(2n^*)]}{(\mathcal{A}^*/n^*)\sinh\mathcal[\mathcal{A}^*/(2n^*)]}\nonumber\\
&\equiv B(\vec z\cdot\vec{\jstatc},\mathcal{A}^*/n^*,\sigma),
\label{eq:mainresult}
\end{align}
where we have used Eq.~\eqref{eq:hdomA} for the third line and
Eq.~\eqref{eq:legendrelambda} for the fifth line.
Although the uniform cycle decomposition was essential in the steps leading
to this result, its final statement is no longer dependent on the specific
choice of independent cycles, as can be seen from the following argument. In
a different complete set of cycles the cycle currents are represented as
$\tilde{\mcJ_\alpha}=\sum_\beta C_{\alpha\beta}\mcJ_\beta$ with some
transformation matrix $C_{\alpha\beta}$. The corresponding scaled cumulant
generating function then reads
\begin{equation}
\tilde\lambda(\tilde{\vec z})=\lim_{t\to\infty}\frac{1}{t}\log\mean{\exp\left[\sum_\alpha\tilde
    z_\alpha\tilde{\mcJ_\alpha}t\right]}=\lim_{t\to\infty}\frac{1}{t}\log\mean{\exp\left[\sum_{\alpha,\beta}\tilde
    z_\alpha
    C_{\alpha\beta}\mcJ_\beta t\right]}=\lambda\left(C^\mathsf{T}\tilde{\vec{z}}\right)
\end{equation}
and satisfies the bound
\begin{equation}
  \tilde\lambda(\tilde{\vec z})\geq
  B(C^\mathsf{T}\tilde{\vec{z}}\cdot\vec{\jstatc},\mathcal{A}^*/n^*,\sigma)=
B(\tilde{\vec{z}}\cdot\tilde{\vec{\jstatc}},\mathcal{A}^*/n^*,\sigma),
\end{equation}
which finally proves Eq.~\eqref{eq:mainresultshort} for arbitrary
cycle decompositions. It should be noted that, \textit{a priori},
$\mathcal{A}^*/n^*$ still refers to the minimal
$\mathcal{A}_\beta/n_\beta$ in a set of uniform cycles. However, in a
possible application, where the network topology and the cycle
affinities are known but individual transition rates are not known,
the more general result~\eqref{eq:mainresultshort}, where
$\mathcal{A}^*/n^*$ refers to the minimum over \emph{all} cycles,
might lead to a weaker bound that is more useful. Since uniform cycles
always have affinity $\mathcal{A}_\beta>0$, the minimization can be
restricted to cycles with positive affinity. Thus, in the physically
important case of a network containing cycles with zero affinity,
Eq.~\eqref{eq:mainresultshort} provides a bound that is stronger than
the parabolic bound that is obtained for $\mathcal{A}^*/n^*\to 0$.
This fact is in accordance with the insight from Ref.~\cite{pugl10},
stating that coarse graining such cycles with zero affinity has little
effect on the fluctuations of the entropy production.

\paragraph{Vanishing currents and equilibrium.} So far, we have
considered only the case of non-vanishing stationary currents
$\jstat_{ij}$ along all edges. If this is not the case for some edges
(although $k_{ij}>0$), we have to split the bound for the large
deviation function in Eq.~\eqref{eq:masterbound} in two parts as
\begin{equation}
    I(\vec
  J)\leq\sum_{i<j|\jstat_{ij}\neq0}\jstat_{ij}\psi(J_{ij}/\jstat_{ij},f_{ij})+\sum_{i<j|\jstat_{ij}=0}\left[J_{ij}\arsinh\frac{J_{ij}}{a_{ij}}-\sqrt{a_{ij}^2+J_{ij}^2}+a_{ij}\right].
\end{equation}
Evaluating this bound along currents $\vec{J}=\xi \vec{\jstat}$, the
second sum vanishes while the first sum leads along the same lines as
before to the bound \eqref{eq:mainresult}. However, in the equilibrium
case, where all stationary currents vanish, this procedure leads
merely to the trivial statement $\lambda(\vec z)\geq 0$. While typical
fluctuations in equilibrium systems can be well described within
linear response theory, rare fluctuations exhibit many features akin
to systems far from equilibrium \cite{raha13}. A non-trivial
bound for these rare fluctuations in equilibrium systems, similar to the one derived in
Ref.~\cite{piet15} for unicyclic networks, is obtained by letting the
affinity $\mathcal{A}_\alpha$ of a single fundamental cycle go to zero
while keeping the affinities of the other fundamental cycles fixed at
zero. Then, none of the cycles can have an affinity smaller than
$\mathcal{A}_\alpha$, so that $\mathcal{A}^*=\mathcal{A}_\alpha$. The
length of the relevant cycle $n^*$ can be bounded by the total number
of states $N\geq n^*$. Due to the Einstein relation \cite{andr07b},
the stationary current $\jstatc_\alpha$ is given for small
$\mathcal{A}_\alpha$ by
\begin{equation}
  \jstatc_\alpha=D_\alpha\mathcal{A}_\alpha+\mathcal{O}(\mathcal{A}_\alpha^2),
\end{equation}
where $D_\alpha$ is the diffusion coefficient
\begin{equation}
  2D_\alpha\equiv\lim_{t\to\infty}t\mean{(\mcJ_\alpha-\jstatc_\alpha)^2}.
\end{equation}
With the entropy production \eqref{eq:smc} reducing to
$\sigma=\jstatc_\alpha\mathcal{A}_\alpha=D_\alpha\mathcal{A}_\alpha^2$,
we thus obtain as bound on the scaled cumulant generating function for
the equilibrium fluctuations of $\mcJ_\alpha$
\begin{equation}
  \lambda_{\alpha}(z)\geq \lim_{\mathcal{A}_\alpha\to
    0}B(zD_\alpha\mathcal{A}_\alpha,\mathcal{A}_\alpha/N,D_\alpha\mathcal{A}_\alpha^2)=2N^2
  D_\alpha(\cosh(z/N)-1).
\label{eq:eqbound}
\end{equation}
This bound is saturated for $z\ll 1$, where $\lambda_\alpha(z)$ depends
quadratically on $z$ according to the Gaussian distribution of typical
fluctuations within linear response. The probability of extreme fluctuations
beyond linear response deviates the more from this Gaussian shape the
smaller the number of states in the network is.

\paragraph{Conclusion.} We have shown that for every Markov network it
is possible to construct a set of independent cycles such that all
affinities and stationary cycle currents are positive and that all
stationary currents along edges are aligned with the direction of the
cycles. We call this a decomposition in \textit{uniform} cycles. Based
on such a decomposition of the stationary currents, we have proven the
hyperbolic cosine shaped bound on the scaled cumulant generating
function for current fluctuations \eqref{eq:mainresultshort}, which we
have previously conjectured in Ref.~\cite{piet15}. This bound refines
the parabolic bound previously proven in
Ref.~\cite{ging16} and requires knowledge of the minimal positive
affinity per number of edges in any cycle of the network. The bound is
universally valid for arbitrary currents in arbitrary networks and
entails a bound on the Fano factor in enzyme kinetics, previously
conjectured in Ref.~\cite{bara15a}, which is thereby proven as well.
We have also proven a new universal bound on equilibrium fluctuations,
given in Eq.~\eqref{eq:eqbound}, that depends only on the diffusion
coefficient and on the number of states.

\begin{appendix}
  \section{Proof of the concavity of $\psi(\zeta,f)$ in $f$}
  \label{sec:appa}
The derivative of $\psi(\zeta,f)$ in Eq.~\eqref{eq:psidef} can be written as
\begin{equation}
  \partial_f\psi(\zeta,f)=\frac{1}{2}\sqrt{b^2+\zeta^2}\sqrt{b^2+1}-\frac{b^2}{2}-\frac{\zeta}{2}.
\end{equation}
As a function of $b=[\sinh(f/2)]^{-1}$ (which decreases
monotonically in $f$), this expression increases
monotonically for $b>0$, since
\begin{equation}
  \partial_b\partial_f\psi(\zeta,f)=\frac{b}{2}\left(\sqrt{\frac{b^2+\zeta^2}{b^2+1}}+\sqrt{\frac{b^2+1}{b^2+\zeta^2}}-2\right)\geq 0
\end{equation}
(note that $x+1/x\geq 2$ for $x>0$). Therefore, $\partial_f\psi$ decreases
monotonically in $f$ and
\begin{equation}
  \partial_f^2\psi(\zeta,f)<0
\end{equation}
holds for all $\zeta\in\mathbb{R}$ and $f>0$.

\section{Proof of the monotonic decrease of $\psi(\zeta,f)/f$ in $f$}
\label{sec:appb}
The monotonic decrease of $\psi(\zeta,f)/f$ in $f$ for $f>0$ is equivalent to the
monotonic increase of its Legendre transform
\begin{equation}
  \mu(z,f)\equiv \max_\zeta[\zeta
  z-\psi(\zeta,f)/f]=\lambda(zf,f)/f=\frac{\cosh[(z+1/2)f]-\cosh(f/2)}{f\sinh(f/2)}\equiv\frac{A(z,f)}{B(f)}.
\end{equation}
The derivative $\partial_f\mu(z,f)$ is non-negative if
\begin{align}
  0\leq C(z)\equiv&\partial_f A(z,f)\,B(f)-A(f)\,\partial_fB(f)\nonumber\\
  &=\left[(z+1/2)\sinh((z+1/2)f)-(1/2)\sinh(f/2)\right]\,f\,\sinh(f/2)\nonumber\\
  &\qquad-\left[\cosh((z+1/2)f)-\cosh(f/2)\right]\,\left[\sinh(f/2)+(f/2)\cosh(f/2)\right].
\end{align}
Equating $\partial_z C(z)$ to zero leads to
\begin{equation}
	2(z+1/2)\tanh(f/2)=\tanh[(z+1/2)f],
\end{equation}
which has the three solutions $z=-1,-1/2,0$. The stationary points $z=0$ and
$z=-1$ are minima since
\begin{equation}
  \partial_z^2 C(z=0)=\frac{f^2}{2}(\sinh f-f)>0
\end{equation}
and $C(z)$ is symmetric with respect to $-1/2$. Thus, the center of symmetry
at $z=-1/2$ is a maximum of $C(z)$ and the global minimum of $C(z)$ is given
by $C(0)=C(-1)=0$.

\end{appendix}

\section*{References}

\bibliographystyle{utphys}
\bibliography{refs}

\end{document}